\documentclass[english,twocolumn]{article}

\usepackage[big]{dgruyter}
\usepackage[utf8]{inputenc}

\usepackage{microtype}
\usepackage{float}
\usepackage{graphicx}
\usepackage{amssymb,amsfonts}
\usepackage{amsmath}

\usepackage{siunitx}
\DeclareSIUnit[number-unit-product=]\percent{\char`\%} 

\renewcommand{\Re}{\mathrm{Re}}
\renewcommand{\Im}{\mathrm{Im}}

\begin{document}

\articletype{Research article}

\author*[1]{E. Galiffi}
\author[1,2]{P. A. Huidobro}
\author[3,4]{P. A. D. Gon\c{c}alves} 
\author[3,4,5]{N. A. Mortensen} 
\author[1]{J. B. Pendry} 
\affil[1]{The Blackett Laboratory, Imperial College London, UK}
\affil[2]{Instituto de Telecomunicações, Instituto Superior Técnico-University of Lisbon, Avenida Rovisco Pais 1,1049-001 Lisboa, Portugal}
\affil[3]{Center for Nano Optics,
University of Southern Denmark, Campusvej 55, DK-5230 Odense M, Denmark}
\affil[4]{Center for Nanostructured Graphene, Technical University of Denmark, DK-2800 Kongens Lyngby, Denmark}
\affil[5]{Danish Institute for Advanced
Study, University of Southern Denmark, Campusvej 55, DK-5230 Odense M, Denmark}

\title{Probing Graphene's Nonlocality with Singular Metasurfaces}
\abstract{Singular graphene metasurfaces, conductivity gratings realized by periodically suppressing the local doping level of a graphene sheet, have recently been proposed to efficiently harvest THz light and couple it to surface plasmons over broad absorption bands, achieving remarkably high field enhancement. However, the large momentum wavevectors thus attained are sensitive to the nonlocal behaviour of the underlying electron liquid. Here, we extend the theory of singular graphene metasurfaces to account for the full nonlocal optical response of graphene and discuss the resulting impact on the plasmon resonance spectrum. Finally, we propose a simple local analogue model that is able to reproduce the effect of nonlocality in local-response calculations by introducing a constant conductivity offset, which could prove a valuable tool in the modelling of more complex experimental graphene-based platforms.}


\maketitle

\section{Introduction}

Over the past two decades, singular plasmonic structures, such as touching metallic wires and spheres, have demonstrated enticing capabilities for controlling light in the subwavelength regime thanks to their ability to bridge very different length scales, namely the wavelength of the photon and that of the electron~\cite{luo2010surface,aubry2010plasmonic,estakhri2013physics}. Characterized by features much smaller than their overall size, these structures have so far enabled extreme confinement of electromagnetic fields, with a plethora of far-reaching applications, including the access to quantum regimes of light--matter interactions~\cite{chikkaraddy2016single,zhu2016quantum,fernandez-domiguez2018}. More recently, extended structures featuring singularities have been investigated in the context of metasurfaces~\cite{galiffi2018broadband,yang2018transformation}, which enable larger scattering cross-sections and lower losses, as well as unprecedented tunability and dynamical control of electromagnetic waves~\cite{kildishev2013planar,yu2014flat,shaltout2019spatiotemporal}.

\begin{figure}[b!]
    \includegraphics[width=\columnwidth]{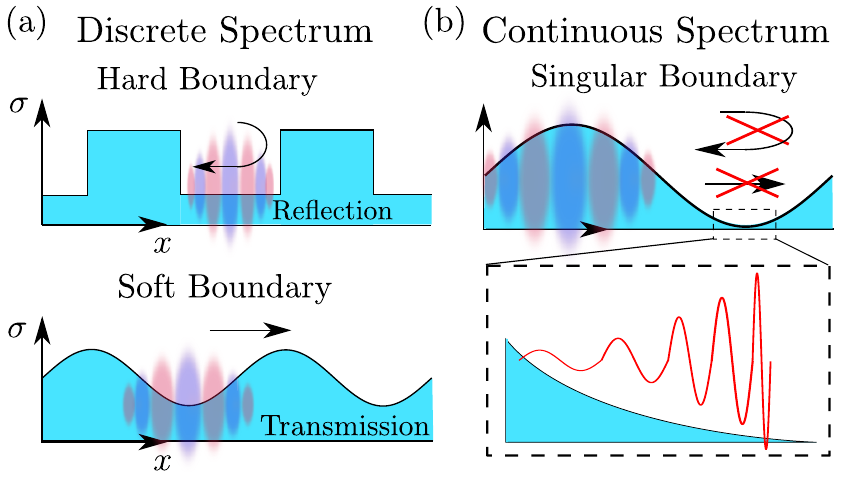}
    \caption{(a) The in-plane scattering of an electromagnetic wave in a periodic system, e.g., a plasmon propagating along a periodically modulated conductive surface is typically dominated by reflection at hard-boundaries or transmission through soft boundaries, leading to discrete Fabry--P{\'e}rot modes or Bloch waves, respectively. (b) At a singular boundary, both transmission and reflection channels are virtually inaccessible, and the only available path for a wave is to shrink its wavelength and concentrate its energy as it travels towards the singular point.}
    \label{fig:singularities}
\end{figure}

The working principle of singular structures, which has been recently shown to be intimately linked to the concept of compactification encountered in high-dimensional field theories~\cite{pendry2017compacted,galiffi2019singular}, may be summarized in the following consideration. In a conventional one-dimensional (1D) periodic scattering problem (Fig.~\ref{fig:singularities}), one can identify two distinct scenarios: hard-boundary scattering, which is often modelled through boundary conditions, commonly results in reflection, and the subsequent quantization of scattered fields into effective Fabry--P{\'e}rot modes (Fig.~\ref{fig:singularities}a); the opposite regime consists of the weak scattering limit, often modelled with WKB-type approaches, whose main effect is the phase change of a largely transmitted wave (Fig.~\ref{fig:singularities}b). Singular structures constitute a narrow intermediate regime, whereby the scattering process is not abrupt enough to generate significant back-reflection, whilst not being smooth enough to let the wave be significantly transmitted. As a result, the wavelength of the excitation becomes increasingly short as it approaches a so-called singular point. Its group velocity is dramatically reduced, such that the wave never reaches the singularity, and energy is absorbed close to it in the presence of material loss, and realising remarkable concentration of electromagnetic energy within nanoscale volumes. Recently, graphene-based singular metasurfaces have been proposed as a promising platform for the focusing of THz plasmons, as well as for their broadband, tunable plasmonic response to far-field illumination~\cite{galiffi2018broadband}. The plasmonic response of graphene has recently demonstrated unprecedented field confinement, concentrating waves which propagate with free-space wavelengths of tens to hundreds of microns down to the atomic scale~\cite{iranzo2018probing,lundeberg2017tuning,dias2018probing,Ni:2018}. In addition, the technological relevance of these THz plasmons for vibrational sensing~\cite{rodrigo2015mid,li2014graphene,gonccalves2016introduction,grigorenko2012graphene,low2014graphene} and high-speed wireless communication~\cite{Vicarelli:2012,poumirol2017electrically,Yang:2017} has attracted enormous interest in these surface excitations.

However, it has recently been shown that the account of nonlocal effects---arising from the quantum nonlocal response of the two-dimensional (2D) electron gas---is of paramount importance when the plasmon wavelength becomes comparable to the electronic Fermi wavelength, in order to correctly predict their electromagnetic response~\cite{lundeberg2017tuning,dias2018probing}. The nonlocal response of singular metallic structures featuring three-dimensional electron gases has been widely studied~\cite{raza2015nonlocal}, primarily via the so-called hydrodynamic model~\cite{boardman1982electromagnetic,PhysRevB.26.7008}, which accounts for charge screening at a dielectric--metal interface~\cite{garcia2008nonlocal,toscano2013nonlocal,yang2019nonlocal}. Alternative theoretical models have also been developed in the past, which simplify the account of nonlocal effects in complex plasmonic structures~\cite{luo2013surface,mortensen2014GNOR,yan2015pdm,christensen2017feibelman}. More recently, nonlocal effects have attracted renewed interest due to their surprising role in the reduction of plasmonic losses~\cite{dias2018probing}, and, in particular, due to the sizable impact of quantum mechanical effects in plasmon-enhanced light--matter interactions~\cite{gonccalves2019plasmon} in the nanoscale, as well as for applications to all-optical signal processing~\cite{kwon2018nonlocal}.

\begin{figure}[t]
    \includegraphics[width=\columnwidth]{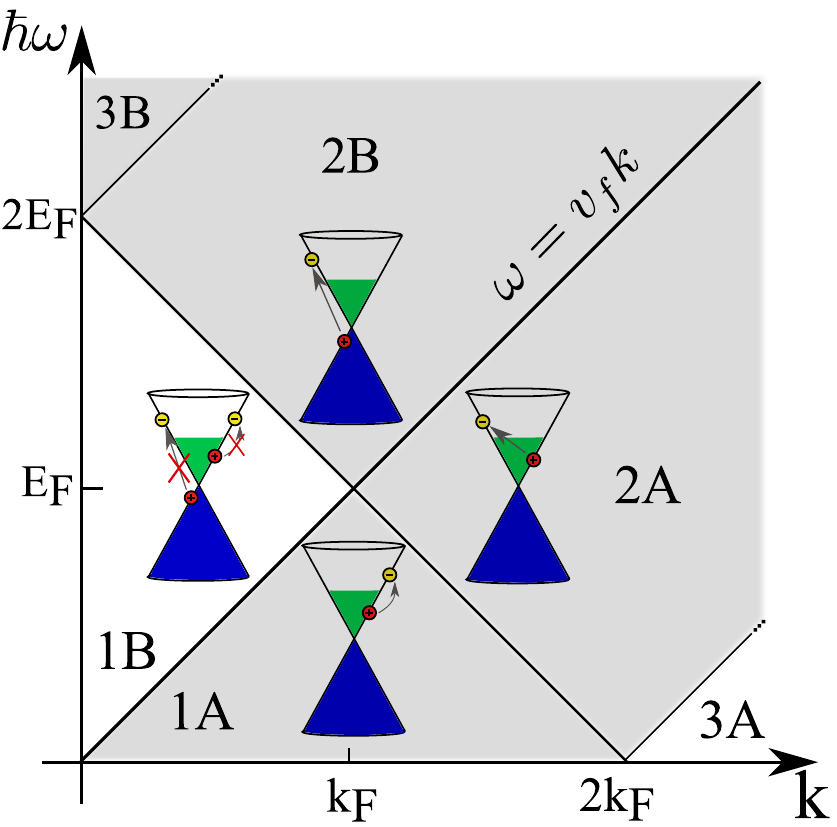}
    \caption{Electronic contributions to the graphene conductivity in different regions of phase space~\cite{gonccalves2016introduction}. Region 1B ($k<\omega/v_F$, $k<2 k_F-\omega/v_F$) of phase space is protected from Landau damping arising from both interband and intraband transitions. The lossy (shaded) regions are: 1A ($\omega/v_F<k< 2k_F-\omega/v_F$) and 2A ($\omega/v_F< k<2 k_F + \omega/v_F$, $k>2k_F-\omega$), dominated by Landau damping resulting from intraband transitions, and 2B ($2 k_F-\omega/v_F<k<\omega/v_F$, $\omega/v_F<2k_F+k$) and 3B ($k<\omega/v_F-2 k_F$) dominated by indirect and direct interband transitions respectively.}
    \label{fig:graphene_nonlocality}
\end{figure}

In these singular metasurfaces, the nonlocal response of graphene arises from the onset of different types of electronic transitions within the regions of phase space shown in Fig~\ref{fig:graphene_nonlocality}. Region 1B constitutes the so-called lossless regime (in the absence of electronic scattering processes). Here, interband transitions are forbidden due to Pauli blocking, and the small plasmon momentum---i.e., $k\ll k_F$, where $k_F$ is the Fermi wavevector---does not allow for any indirect transitions. Hence, in this regime, the only loss channels for graphene plasmons arise from electronic scattering processes (e.g., with phonons, defects, etc)~\cite{Yan:2013,Ni:2018} which are commonly introduced phenomenologically via the so-called relaxation-time approximation~\cite{gonccalves2016introduction}. Nevertheless, the incorporation of quantum nonlocal effects is reflected in the reactive (imaginary) component of the conductivity for large plasmon momenta $k\to \omega/v_F$. In fact, the divergent character of graphene's conductivity at the boundary between region 1B and 1A constitutes a main detrimental effect for the realization of conductivity singularities in graphene. Region 1A suffers from the onset of Landau damping, which arises due to the matching between the phase velocities of the electrons and of the plasmons. This has the effect of dramatically enhancing the loss. Similarly, region 2A is affected by additional intraband channels, which become accessible once the plasmon momentum $k>k_F$. Finally, indirect (region 2B) and direct (region 3B)  interband transitions occur once the plasmon energy $\hbar \omega > 2E_F-\hbar v_F k$ and $\hbar \omega > 2E_F+\hbar v_F k$ respectively.

\begin{figure*}[t]
    \includegraphics[width=\textwidth]{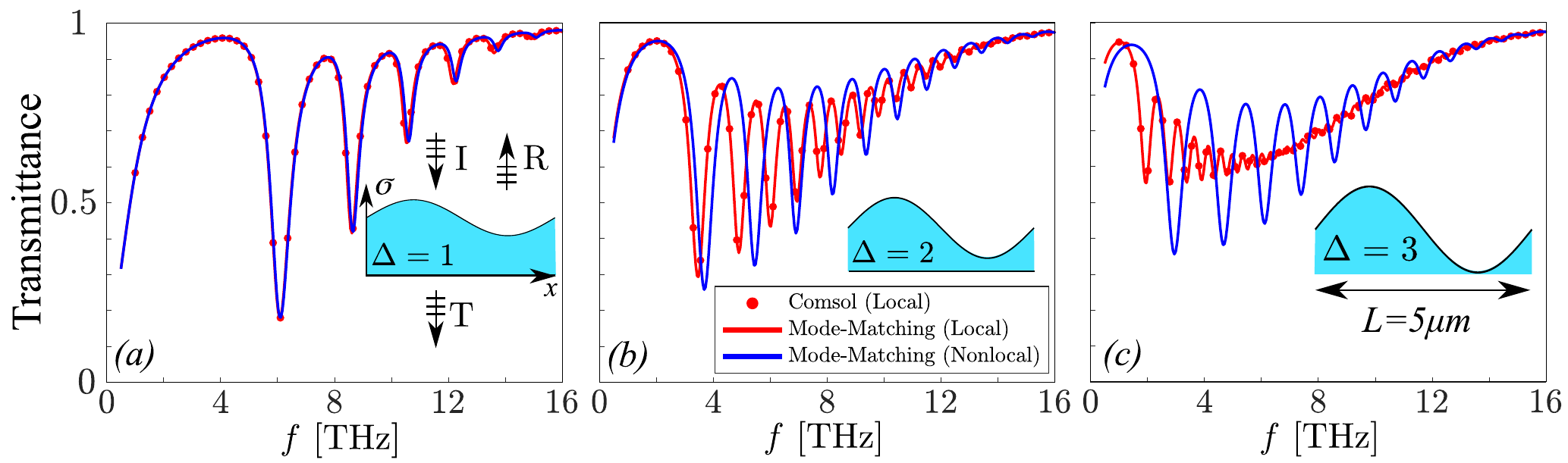}
    \caption{Local (red) and nonlocal (blue) transmittance spectra for plane wave illumination through the graphene metasurface at normal incidence, obtained with the mode-matching (continuous lines) and finite-element method (dots) for three increasingly singular metasurfaces corresponding to $\Delta = 1$ (a), $\Delta = 2$ (b) and $\Delta = 3$ respectively. The nonlocal contribution, which is negligible away from the singular regime, becomes dominant as the singular limit is approached, opposing the merging of surface plasmon modes. }
    \label{fig:transmission}
\end{figure*}

Due to the extreme values of plasmon momenta to which a singular structure can couple incident photons to, a rigorous account of the momentum-dependence of the optical response of these metasurfaces is pivotal. In this work, we explore the nonlocal behaviour of plasmons in singular graphene metasurfaces and show that these systems are able to probe the strong nonlocal response of 2D electron gases by coupling far-field radiation to deeply subwavelength plasmon modes. By means of a nonlocal mode-matching technique~\cite{gonccalves2016introduction}, supported by numerical calculations, as well as a phenomenological local-analogue model, we unravel the physics underpinning the onset of nonlocality in these metasurfaces. We believe that our method constitutes a valuable tool for incorporating nonlocal effects in complex metasurface setups, and may be employed as an alternative approach to fully nonlocal conductivity models.

\section{Methods}

Nonlocal effects in plasmonics manifest themselves when the plasmon wavelength approaches the typical electronic wavelength $\lambda_F$ in a material. In this regime, the spatial variation of the electric field $\mathbf{E}(x)$ is sufficiently abrupt to sample the underlying inhomogeneity of the electron gas, so that the constitutive relation for the surface current density can be written as
\begin{align}
    J(x,\omega) = \int \sigma(x-x',\omega) E_x(x',\omega) dx' 
\end{align}
and thus can no longer be approximated assuming a spatial dependence of the conductivity of the form $\sigma(x-x',\omega) = \sigma(\omega)\delta(x-x')$, where $\delta(x)$ is the Dirac delta function.

However, when the structuring of a THz metasurface is performed over scales much larger than the Fermi's wavelenth ($L \gg \lambda_F$), a separation of length scales can be assumed. Hence, we can write, under the adiabatic approximation:
\begin{align}
    J(x,\omega) = \int \sigma(x-x',\omega)\zeta(x') E_x(x',\omega) dx'
\end{align}
where $\zeta(x')$ is a dimensionless variable which describes the spatial modulation of the conductivity of graphene~\cite{slipchenko2013analytical,huidobro2016graphene}, the latter depending monotonically on the \emph{local} doping level of graphene. This has the desirable property of being actively tunable (e.g., electrostatically, chemically, or optically). In this work, we assume that a periodic conductivity modulation is applied, which, for simplicity and definiteness, is herein assumed to be of the form $\zeta(x) = 1 + \zeta_1\cos(gx)$, where $L= 2\pi/g$ is the period of the 1D metasurface and $g$ the reciprocal lattice vector associated with the same.

Using Bloch's theorem and expanding the Bloch modes of the in-plane electric field and the surface current as a Fourier series, one may write
\begin{align}
    E_x(x) = e^{ikx}\sum_n E_{n,x} e^{ingx}
\end{align}
and a simple relation between the Fourier amplitudes of the electric field and the surface current hereby takes the form
\begin{align}
    J_{n,x} = \sigma(k+ng)[E_{n,x} + \frac{\zeta_1}{2}(E_{n+1,x}+E_{n-1,x})]
\end{align}
which is accurate as long as the reciprocal lattice vector of the metasurface satisfies $g\ll k_F$. For concreteness, the nonlocal conductivity model~\cite{gonccalves2016introduction} is described in Appendix~\ref{sec:nonlocal_conductivity_appendix}.

\section{Results}

\begin{figure*}[t]
    \includegraphics[width=\textwidth]{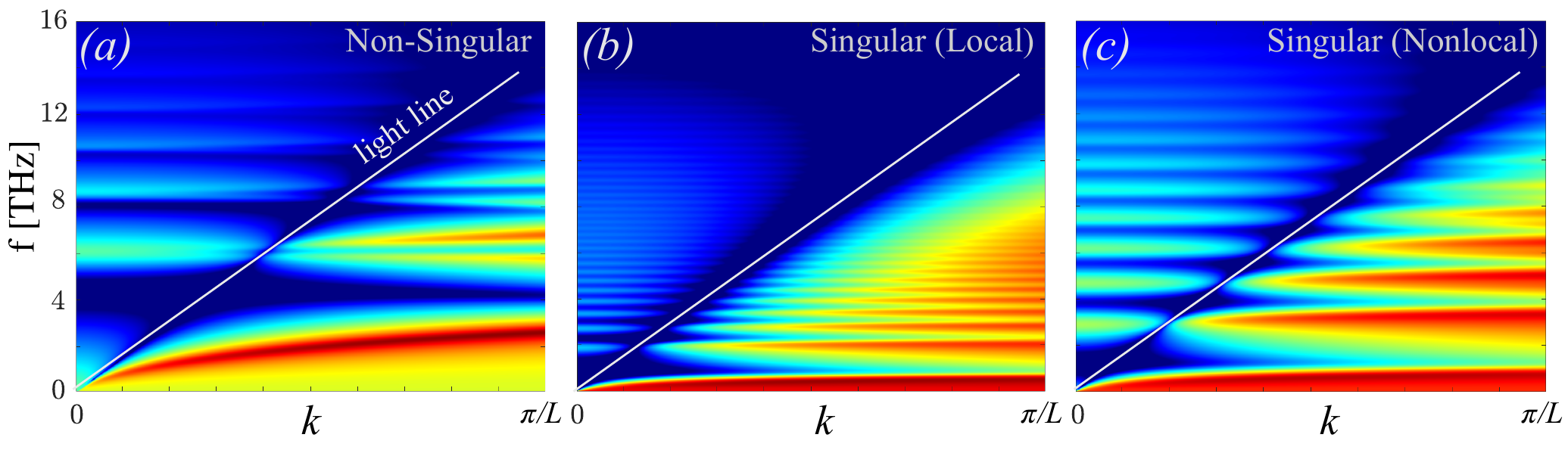}
    \caption{Band structure for $\Delta = 1$ (a) and $\Delta = 3$ (b,c), visualized by plotting the logarithm of the absolute value of the reflection coefficient. Local (a,b) and nonlocal (c) spectra differ significantly for the singular case only. Moreover, in the singular limit, it can be seen that modes above each band gap become extremely broad, due to their stronger radiative coupling. In addition, the modes are effectively degenerate at $k=0$, due to the symmetry of the modulation combined with the strongly quasi-static character of graphene plasmons.}
    \label{fig:band_structure}
\end{figure*}

The main effect of nonlocality in graphene is to oppose the formation of a singularity by increasing the conductivity probed by large-momentum Fourier components. In Fig.~\ref{fig:transmission} we plot the transmission spectra under plane wave illumination at normal incidence ($k=0$) for different modulation strengths $\Delta = -\log_{10}{(1-\zeta_1)}$, corresponding to the number of orders of magnitude by which the conductivity is suppressed at the singular point. We assume an average Fermi level $E_F = \SI{0.4}{\eV}$, a conductivity grating period $L= \SI{5}{\um}$, and a mobility $\mu_\mathrm{m} = \SI[per-mode=symbol]{e4}{\cm\squared \per \volt \per \second }$ resulting in an electron scattering time $\tau = \mu_\mathrm{m} E_F/(v_F^2 e) \approx \SI{0.44}{\pico\second}$, where the Fermi velocity $v_F = \SI{9.5e5}{\meter\per\second}$ \cite{fan2015tunable} is assumed. Our results are obtained via the nonlocal mode-matching method outlined above; these have been benchmarked, in the local-response limit, against finite-element method (FEM) numerical calculations using a commercially available package (COMSOL Multiphysics).
For weak conductivity modulation, i.e., far from the singular limit (Fig.~\ref{fig:transmission}a), the local and nonlocal spectra are effectively equivalent. In this limit, only momentum states well below the Landau damping regime $k \approx \omega/v_F$ are populated, so that the metasurface can be accurately described via a local Drude-type conductivity model $\sigma_D (\omega) = \frac{e^2}{\pi \hbar^2} \frac{E_F}{(\gamma-i\omega)}$, where $\gamma = \tau^{-1}$. As we increase the modulation strength to 99.9\% of the average value (Fig.~\ref{fig:transmission}b, $\Delta = 2$) the local and the nonlocal spectra start deviating, the latter exhibiting a clear blueshift which is consequence of nonlocality (see, e.g., Ref.~\cite{gonccalves2016introduction}), since plasmon resonance frequencies $\omega \propto \sigma$ (see dispersion relation, Eq.~\ref{eq:disp_rel}). Finally, for $\Delta = 3$ (Fig.~\ref{fig:transmission}), nonlocality becomes a dominant effect, which effectively saturates the plasmonic spectrum, opposing any further merging of the plasmon resonances.

Away from $k=0$, additional effects are present, as shown in Fig.~\ref{fig:band_structure}, where we plot in log-scale the absolute value of the reflection coefficient, which has been colour-saturated in order to allow both propagating and evanescent modes to be identifiable. In the non-singular regime [Panel~ (a)], plasmonic band gaps are clearly visible at $k = \pi/L$, whereas no significant gaps are present at $k=0$, due to the quasistatic character of these excitations, as discussed in Ref.~\cite{kraft2015designing}. However, as the singular limit is approached, the band gap becomes so large that the two resonances become indistinguishable due to their finite width, resulting in extremely flat bands, an effect which survives the onset of nonlocality [Panel~(c)]. In this regime, the overall effect of nonlocality is not only to oppose the merging of the resonances, but also to introduce significant additional broadening due to nonlocal intraband Landau damping.

The account of nonlocality can be somewhat demanding in the modelling of more complex experimental setups. Consequently, local-analogue models which are able to incorporate the effects of nonlocality in a local simulation are valuable tools for the theoretical modelling of plasmonic systems. Here we propose a simple local-analogue model which can accurately reproduce the results of the fully nonlocal calculation carried out above. Local-analogue models were originally proposed for metallic plasmonic systems~\cite{luo2013surface} in order to capture nonlocal effects under the framework of the hydrodynamic model of the free-electron gas at the interface between nearly-touching metallic structures. In that context, the effect of nonlocality is the inward shift of the induced charges, i.e., away from the metallic surface and into the bulk, thereby effectively widening the gap between the components of the dimer (e.g., metallic cylinders or spheres). Consequently, the substitution of a thin metallic layer by \emph{an effective} dielectric one was able to accurately reproduce the optical response of such nearly-touching metallic structures.

\begin{figure}[t]
    \includegraphics[width=\columnwidth]{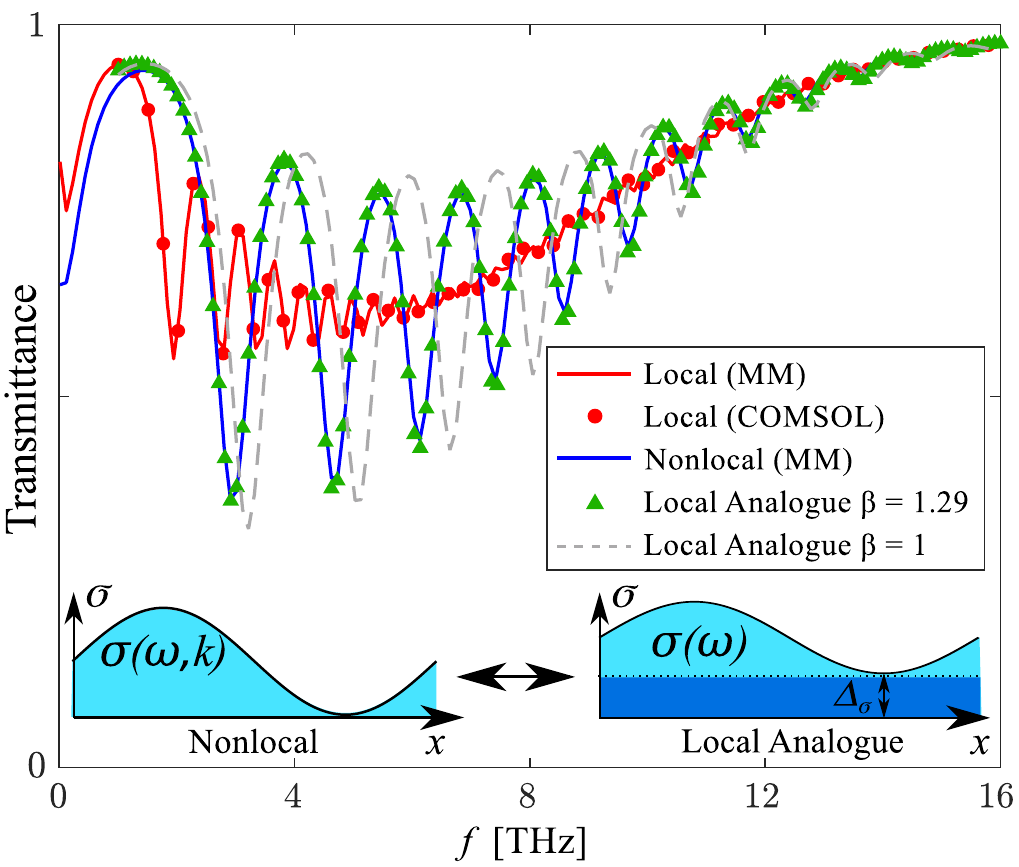}
    \caption{Local (red), nonlocal (blue line) and local analogue (green triangles for $\beta=1.29$ and grey dashed line for $\beta=1$) transmittance spectra of the singular ($\Delta = 3$) graphene conductivity grating. The inset shows how a local analogue metasurface can be obtained by saturating the conductivity of graphene near the value which causes the local plasmon dispersion to cross the electron dispersion $\omega = v_F k$, a regime dominated by Landau damping.}
    \label{fig:local_analogue_model}
\end{figure}

Conversely, the type of singular structure described in this work entails the inverse effect: since the conductivity is strongly enhanced as $k\to \omega/v_F$, the effect of nonlocality is to smear out the singularity by effectively saturating the local conductivity to a minimum level $\sigma_{s}$ dictated, qualitatively, by the condition $k(\sigma_{s}) \approx \omega/v_F$, i.e., when the plasmon wavelength $\lambda_\mathrm{p} \to \lambda_F$, and Landau damping opposes any further confinement of the plasmonic field. The quasi-static dispersion relation of graphene plasmons is reads~\cite{gonccalves2016introduction}:
\begin{align}
    \epsilon_1 + \epsilon_2 + i\frac{\sigma}{\epsilon_0 \omega}k = 0,
    \label{eq:disp_rel}
\end{align}
where $\sigma \equiv \sigma(k,\omega)$ and $\sigma \equiv \sigma(\omega) = \sigma(k \to 0,\omega)$ in the nonlocal and local cases, respectively. Herein, we have set $\epsilon_{1,2} = 1$ (for simplicity alone). Moreover, we can then substitute the wavevector $k = \beta \omega/v_F$, where $\beta$ is a phenomenological factor of order $\sim 1$ which quantifies the fraction of electron momentum to which the plasmon can couple before saturating (which is exactly one if momentum saturation occurs exactly at the electron momentum). In this fashion, we thus obtain the saturation value for the conductivity,  $\sigma_{s} = 2i  \epsilon_0 v_F / \beta $. In Fig.~\ref{fig:local_analogue_model}, we add a positive surface conductivity offset
\begin{align}
    \Delta_{\sigma}(\omega) = i\Im{[\sigma_s - (1-\zeta_1)\sigma_D(\omega)]} [1-i/(\omega \tau)]
\end{align}
in a local FEM calculation, where the factor in the first square bracket is responsible for the smearing of the imaginary part of the surface conductivity, whereas the second ensures that the loss-tangent $\Re{[\sigma]}/\Im{[\sigma]}$ is preserved upon the conductivity offset.

For $\beta = 1$, the agreement between the previous nonlocal result (Fig.~\ref{fig:transmission}c) and the spectrum obtained using the local analogue model is only qualitative. However, as the figure plainly shows, by choosing $\beta \simeq 1.29$ this simple model is able to reproduce the entire transmission spectrum with remarkable accuracy, hereby validating the physical assumptions behind our local analogue model, and providing us with a useful and intuitive method for the incorporation of nonlocal effects in the future modelling of complex metasurfaces based on 2D materials.

\section{Conclusions}

In this work we have presented a theoretical description of nonlocal effects in singular graphene metasurfaces. By calculating the transmission spectra under plane wave illumination, as well as the plasmon band structure, we have demonstrated how such conductivity gratings are able to probe the nonlocal response of graphene. Furthermore, we have discussed the consequent limitations imposed by nonlocality to the field confinement and spectral degeneracy induced by the singularity, which is effectively smeared out by the increased conductivity probed by large plasmon wavevectors. Finally, we have proposed a simple local-analogue model which is able to reproduce the effects of nonlocality by means of an effective surface conductivity offset, which saturates the plasmon wavevector to the electronic one. To conclude, singular graphene metasurfaces constitute a platform for probing nonlocality in graphene with far field measurements. Our results form the the basis for a quantitative account of nonlocality in these metasurfaces, and should be valuable for guiding future experimental efforts.

\begin{acknowledgement}
We would like to thank A.~I. Fern\'{a}ndez-Dom\'{i}nguez and F.  Koppens for useful discussions. E.~G. was supported through a studentship in the Centre for Doctoral Training on Theory and Simulation of Materials at Imperial College London funded by the EPSRC (EP/L015579/1). P.~A.~H. acknowledges funding from Funda\c{c}\~ao para a Ci\^encia e a Tecnologia and Instituto de Telecomunica\c c\~oes under project CEECIND/03866/2017. J.~B.~P. acknowledges funding from the Gordon and Betty Moore Foundation. 
N.~A.~M. is a VILLUM Investigator supported by VILLUM FONDEN (grant No. 16498). The Center for
Nano Optics is financially supported by the University of Southern Denmark (SDU 2020 funding). The Center for Nanostructured Graphene is sponsored by the Danish National Research Foundation
(Project No. DNRF103).

\end{acknowledgement}

\appendix
\section{Nonlocal conductivity model}
\label{sec:nonlocal_conductivity_appendix}

The nonlocal conductivity of graphene can be written in terms of graphene's 2D polarizability as~\cite{gonccalves2016introduction}
\begin{align}
    \sigma(k,\omega) = ie^2 \frac{\omega}{k^2}P_{\gamma}(k,\omega)
\end{align}
where $P_{\gamma}(k,\omega)$ is the 2D density-density response function (or 2D polarizability) in the relaxation-time approximation (which incorporates a finite plasmon lifetime whilst preserving electron number density \cite{Mermin:1970,gonccalves2016introduction}). The 2D polarizability in the relaxation-time approximation is given by~\cite{Mermin:1970,gonccalves2016introduction}
\begin{align}
    P_{\gamma}(k,\omega) = \frac{(1+i\gamma /\omega)P(k,\omega + i\gamma)}{1+i\gamma/\omega \cdot P(k,\omega+i\gamma)/P(k,0)}
\end{align}
where $P(k,\omega)$ denotes the zero-temperature density-density response function  in the four regions outlined in Fig.~\ref{fig:graphene_nonlocality} may be written as:

\[ 
 \Re{[P]} = \left\{
\begin{array}{ll}
      -F + \frac{F}{8} \frac{\Bar{k}^2}{\sqrt{\Bar{\omega}^2-\Bar{k}^2}}[ C_h( \frac{\Bar{\omega}+2}{\Bar{k}}) - C_h( \frac{2-\Bar{\omega}}{\Bar{k}}) ], & 1B \\
      -F, & 1A \\
      -F + \frac{F}{8} \frac{\Bar{k}^2}{\sqrt{\Bar{\omega}^2-\Bar{k}^2}}C_h( \frac{\Bar{\omega}+2}{\Bar{k}}), & 2B \\
      -F + \frac{F}{8} \frac{\Bar{k}^2}{\sqrt{\Bar{k}^2-\Bar{\omega}^2}}C( \frac{2-\Bar{\omega}}{\Bar{k}}), & 2A \\
      -F + \frac{F}{8} \frac{\Bar{k}^2}{\sqrt{\Bar{\omega}^2-\Bar{k}^2}}[C_h( \frac{\Bar{\omega}+2}{\Bar{k}})-C_h(\frac{\Bar{\omega}-2}{\Bar{k}})], & 3B \\
      -F + \frac{F}{8} \frac{\Bar{k}^2}{\sqrt{\Bar{k}^2-\Bar{\omega}^2}}[C( \frac{\Bar{\omega}+2}{\Bar{k}})+C(\frac{2-\Bar{\omega}}{\Bar{k}})], & 3A
\end{array} 
\right. 
\]

\[ 
 \Im{[P]} = \left\{
\begin{array}{ll}
      0, & 1B \\
      \frac{F}{8} \frac{\Bar{k}^2}{\sqrt{\Bar{k}^2-\Bar{\omega}^2}}[ C_h( \frac{2-\Bar{\omega}}{\Bar{k}}) - C_h( \frac{\Bar{\omega}+2}{\Bar{k}}) ], & 1A \\
      \frac{F}{8} \frac{\Bar{k}^2}{\sqrt{\Bar{\omega}^2-\Bar{k}^2}}C( \frac{\Bar{\omega}+2}{\Bar{k}}), & 2B \\
      -\frac{F}{8} \frac{\Bar{k}^2}{\sqrt{\Bar{k}^2-\Bar{\omega}^2}}C_h( \frac{2-\Bar{\omega}}{\Bar{k}}), & 2A\\
      -\frac{F}{8}\frac{\Bar{k}^2}{\sqrt{\Bar{\omega}^2-\Bar{k}^2}}, & 3B\\
      0, & 3A
\end{array} 
\right. 
\]
\noindent where $\Bar{k} = k/k_F$, $\Bar{\omega} = \hbar\omega/E_F$, the constant $F = \frac{2k_F}{\pi\hbar v_F}$ and the auxiliary functions: 

\begin{align}
    C_h(z) &= z\sqrt{z^2-1} - \cosh^{-1}{(z)}, \nonumber \\ 
    C(z) &= z \sqrt{1-z^2} - \cos^{-1}{(z)}. \nonumber
\end{align}{}

\bibliographystyle{unsrt}
\bibliography{nonlocalitySingularGraphene}


\end{document}